**Step-directed Epitaxy of Uni-directional Hexagonal Boron Nitride on Vicinal Ge(110)**

*Ju-Hyun Jung[†], Chao Zhao[†], Seong-Jun Yang, Jun-Ho Park, Woo-Ju Lee, Su-Beom Song, Jonghwan Kim, Chan-Cuk Hwang, Seung-Hwa Baek, Feng Ding\*, Cheol-Joo Kim\**

[†]J.-H. Jung, C. Zhao contributed equally to this work.

J.-H. Jung, S.-J. Yang, J.-H. Park, W.-J. Lee, S.-B. Song, J. Kim, S.-H. Baek, C.-J. Kim
Center for Van der Waals Quantum Solids
Institute for Basic Science (IBS)
Pohang University of Science and Technology
Pohang 37673, Republic of Korea
E-mail: kimcj@postech.ac.kr

J.-H. Jung, J.-H. Park, W.-J. Lee, S.-H. Baek, C.-J. Kim
Department of Chemical Engineering
Pohang University of Science and Technology
Pohang 37673, Republic of Korea

C.-J. Kim
Institute for Convergence Research and Education in Advanced Technology
Yonsei University
Seoul 03722, Republic of Korea

C. Zhao, F. Ding
Shenzhen Institute of Advanced Technology
Chinese Academy of Sciences
Shenzhen 518055, China
E-mail: f.ding@siat.ac.cn

C. Zhao, F. Ding
Faculty of Materials Science and Energy Engineering
Shenzhen University of Advanced Technology

Shenzhen 518055, China

S.-B. Song, J. Kim
Department of Materials Science & Engineering
Pohang University of Science and Technology
Pohang 37673, Republic of Korea

C.-C. Hwang
Beamline Research Division
Pohang Accelerator Laboratory
Pohang 37673, Republic of Korea

Insulating hexagonal boron nitride (hBN) films with precisely controlled thickness are ideal dielectric components to modulate various interfaces in electronic devices. To achieve this, high-quality hBN with controlled atomic configurations must be able to form pristine interfaces with various materials in devices. However, previously reported large-scale hBN films with uniform thickness either are polycrystalline or are not suitable for atomically clean assembly via mechanical exfoliation, limiting their applications in device technology. Here, we report the large-scale growth of monolayer single crystalline hBN films on Ge(110) substrates by using chemical vapor deposition (CVD). Vicinal Ge(110) substrates are used for the step-directed epitaxial growth of hBN, where Ge atomic steps act as the hBN nucleation sites, guiding the uni-directional alignments of multiple hBN domains. Density functional theory (DFT) calculations reveal that the optimum hydrogen passivations on both hBN edges and Ge surfaces enable the epitaxial coupling between hBN and the Ge step edges and the single crystallinity of the final hBN films. Using epitaxially grown monolayer hBN films, we fabricate a few hBN films with controlled stacking orders and pristine interfaces through a layer-by-layer assembly process. These films function as high-quality dielectrics to enhance carrier transport in graphene and MoS$_2$ channels.

# 1. Introduction

Hexagonal boron nitride (hBN), whose surfaces are dangling-bond-free, can serve as an ideal passive buffer layer for high-performance electronic and optoelectronic devices by suppressing unwanted interactions between defect states in the insulator and excited charges in the active channels. The inertness of hBN surfaces enables the formation of pristine van der Waals (vdW) interfaces between hBN and arbitrary materials, thereby improving the device performance greatly.[1-3] Additionally, the thickness of the hBN buffer layer can be precisely modulated at the atomic-scale to act as an effective spacer to modulate the interactions between electronic states of constituent materials and the interface. Thereby, advanced devices and material platforms can be realized with desired functionalities, including metallic contacts on two-dimensional (2D) semiconductors with low contact resistance,[4] Josephson junctions for superconducting circuits with low loss,[5] resonant tunnel devices with high operation frequency,[6] and excitonic insulators for exciton circuitry.[7]

While single-crystalline bulk hBN flakes are mostly used for these purposes, the limited scale of bulk crystals, typically below a few tens of micrometers wide, hampers their widespread application for large-scale arrays of devices. hBN films grown by chemical vapor deposition are ideal for large-scale applications, but previous studies have shown either the existence of grain boundaries in the hBN films or interfacial contaminants during the integration process with other active channel materials. These grain boundaries and interfacial contaminants have detrimental effects as trap states on the performance of devices. For instance, the charge carrier mobilities of active channels can be reduced in field-effect transistors due to scattering at the trap states, and the quantum yield of light-emitting materials can be lowered due to trap-assisted recombination. The suppression of defect states in hBN becomes even more critical in device applications where electron-electron interactions or electrical currents occur directly across the hBN for device functionalities.[4-7]

The seamless stitching of uni-directionally aligned hBN domains grown on a wafer-scale single crystalline metal foil, such as Cu[8-11] and Ni[12, 13], by chemical vapor deposition (CVD) approach has been a prevailing method for producing uniform single crystalline hBN films. As revealed in previous computational studies, a zigzag edge of hBN tends to be aligned along a high symmetric direction of the substrate, which can ensure the epitaxial behavior of hBN growth on a single crystal. If the hBN lattice owns all the symmetric operations of the single crystalline substrate, the crystalline alignment of a hBN domain grown on the substrate cannot be changed by any symmetric operation of the substrate[13] and, thus, all the epitaxial hBN islands will be parallelly aligned and finally merged into a large-single crystalline hBN film by seamless stitching. Both experimental[8-14] and theoretical[15] evidence suggests that these uni-directionally aligned hBN grains on an ultra-flat substrate can be seamlessly stitched together, suppressing the formation of tilt and twin grain boundaries in the hBN film. [16,17]

However, hBN films adhered to a metal surface cannot be dry-transferred for clean integration with other device components due to their relatively strong adhesion to metal substrates. This often results in residual contaminants at vdW interfaces after integration, which leads to the in-gap states and the variation of the thickness of hBN spacer. In contrast, hBN films grown on germanium (Ge) surfaces by CVD can be easily dry-exfoliated, allowing for atomically clean assembly.[18] Yet, the uni-directionally aligned hBN grains on Ge surfaces have not been realized because of the limited understanding of the mechanism of hBN growth on Ge.[19] This hinders the clean integration of hBN with a controlled atomic-scale thickness and minimal defective states. We have included a summary table of key figures of merit for the assembly of hBN from various reports in Table S1.

In this study, we report the epitaxial growth of uni-directionally aligned monolayer (ML) hBN islands and the seamless stitching of these islands into a single crystalline hBN film on the vicinal Ge(110) surfaces for atomically clean assembly by CVD (**Figure 1**a). The Ge surface was miscut from the perfect (110) plane by a specific degree to form parallel atomic steps with controllable step width, which reduces the symmetry of the substrate from the $C_{2V}$ of the (110) plane to the $C_s$ or $C_V$. The yield of aligned hBN approaches its maximum when the step edge direction of the vicinal surface is parallel to the polar direction of the hBN lattice. Interestingly, the yield of parallel hBN domains changes non-monotonically when increasing the hydrogen pressure, reaching nearly 100% at an optimized pressure. Our density functional theory (DFT) calculations suggest that hydrogen passivation of both the hBN edge and the substrate is vital for the determination of the hBN orientation on the Ge substrate, demonstrating the importance of the chemical environment for step-directed epitaxial growth of hBN.

**2. Results and Discussion**

**2.1. Epitaxial relationship between hBN and atomic steps of Ge**

Although, in principle, the formation of atomic steps of the Ge(110) surface can eliminate the inversion symmetry of the substrate and facilitate the uni-directional alignments of grown hBN on the substrate, the most efficient atomic steps for efficient hBN epitaxy have not been identified yet. To investigate the effect of step edge orientation on hBN epitaxy, we grew hBN on two different vicinal Ge(110) substrates, whose miscut directions are toward the [001] and [1$\bar{1}$1] axes, and we name them as Sub(A) and Sub(B), respectively (Figure 1b, c). X-ray diffraction (XRD) rocking curve measurements using omega ($\omega$) scans (Figure S1) show that the (110) peaks of the miscut substrates are shifted, confirming that the nominal surfaces have a tilt angle, $\theta_m$, with respect to the [110] crystalline orientation (Figure 1d). Atomic force microscopy (AFM) height images reveal that the atomic steps are oriented perpendicular to the miscut direction (Figure S2). The steps orientations for Sub(A) and Sub(B) are perpendicular to the [1$\bar{1}$0] and [1$\bar{1}\bar{2}$] directions of the substrate, respectively.

While the presence of boron (B) and nitrogen (N) elements and B-N chemical bonds were confirmed by X-ray photoelectron spectroscopy (XPS) and Fourier transform infrared spectroscopy (FTIR) on both Sub(A) and Sub(B) (Figure S3), the grown hBN grains on Sub(A) and Sub(B) show distinct crystallographic alignments. On a vicinal Ge surface with partially covered hBN grains, the atomic steps beneath the triangular-shaped hBN grains remain intact, as can be clearly seen in the AFM phase images (Figure 1e, f). Notably, on either substrate, each triangular hBN grain has an edge along the step direction. On Sub(A), the apex of most triangular hBN grains predominantly points towards the miscut direction, while, on Sub(B), they point either parallel or anti-parallel to the miscut direction with similar ratio (see Figure S4 for scanning electron microscopy (SEM) images and the statistical distribution of grain orientations). Low-energy electron diffraction (LEED) measurements (Figure 1g, h) reveal a single hexagonal pattern of hBN, highlighted in a sapphire blue circle, that shares the same orientation as the pattern of the underlying Ge substrate, highlighted in a dotted red circle. This indicates that all the hBN grains have one zigzag edge along the [1$\bar{1}$0] direction of Ge substrates on both substrates. The six diffraction spots of hBN on Sub(B) have nearly identical intensities, indicating the formation of both anti-parallel hBN grains on Sub(B); In contrast, the pattern of hBN on Sub(A) is clearly asymmetric, suggesting that hBN grains on Sub(A) tends to align along one direction. (Figure 1i, j)

Considering that the interactions between the Ge steps and nucleated hBN edges contribute to the uni-directional growth of hBN by breaking the surface symmetry, the results in Figure 1 indicate that the interaction strength is stronger on Sub(A) than on Sub(B). It is known that Ge forms strong bonds with nitrogen,[19] leading the edges of hBN grains with dangling bonds to form Ge-N bonds to minimize their total free energy. LEED patterns on both Sub(A) and Sub(B) indicate that the zigzag edges of hBN are aligned with the [1$\bar{1}$0] direction of Ge (Figure S5) AFM phase images reveal that all the vertices of triangular hBN grains on Sub(A) have interior angles of 60°, with one edge parallel to the Ge atomic steps in the [1$\bar{1}$0] direction (Figure S5b), indicating that all the hBN edges are N-terminated zigzag edges (Figure 1i). On the other hand, while the hBN grains on Sub(B) have one edge parallel to the [1$\bar{1}$0] direction of Ge, only one vertex has an interior angle of 60° (Figure S5c), resulting in a non N-terminated zigzag edge, aligned to the Ge atomic steps (Figure 1j). Geometric analysis (Figure S6) indicates that the bond strength between the hBN edges and the Ge steps is indeed greater on Sub(A) than on Sub(B), due to the higher density of Ge atoms with dangling bonds along the step (Sub(A): 0.25 nm$^{-1}$, Sub(B): 0.21 nm$^{-1}$). This is consistent with our observations, and the prediction is further confirmed quantitatively with subsequent simulations.

## 2.2. Effect of hydrogen partial pressure on the crystallographic alignment of hBN

We studied how the crystallographic orientations of hBN depend on growth conditions, particularly the hydrogen partial pressure ($P_{H2}$), during the CVD process. On the same surface of Sub(A), hBN crystals were partially grown by modulating $P_{H2}$, while the total growth pressure was kept constant at 760 Torr with the addition of argon flows. Most of the hBN grains form regular triangles with one edge parallel to the Ge step edge. However, the apex between the other two edges points either downward along the Ge terrace, parallel to the miscut direction, or upward on the terrace, anti-parallel to the miscut direction. These are marked as points of red and sapphire color in the schematic and SEM images in **Figure 2**a and b, respectively. The probability of the former type, denoted as $\chi_{align}$, increases with $P_{H2}$, reaching nearly unity at $P_{H2}$ = 0.2, and then decreases again (refer to Figure S7 for the probability distribution as a function of $P_{H2}$). Considering the same type of edge configuration for the triangular grains, $\chi_{align}$ is correlated with the crystallographic alignment of hBN. Indeed, LEED patterns from samples with $\chi_{align}$ values of approximately 0.5 and 1 exhibit 6-fold and 3-fold symmetrical intensity plots, respectively (see Figure 2c and d). This implies that there is growth of hBN with uni-directionally aligned crystallography in the sample with $\chi_{align} \sim 1$, at the optimum $P_{H2}$ value, denoted as $P_{opt,H2}$.[9]

## 2.3. Wafer-scale formations of single-crystalline hBN

The formation of continuous hBN films by merging uni-directionally aligned grains can be scalable over wafer-scale. To observe the formation process of the single-crystalline hBN film, we conducted SEM imaging on as-grown samples on Ge(110) substrates with a miscut toward [001] after different growth times of 4, 8, and 12 hours under the optimal growth condition (**Figure 3**a, b). The image contrast between the bare Ge and hBN-covered regions was significant, allowing us to clearly observe the merging of uni-directionally aligned triangular grains into a continuous film over time. At a surface coverage exceeding 80% (Figure 3b), triangular voids aligned in the same direction were formed, indicating that the crystallographic orientations of the hBN grains were maintained throughout the entire film formation process. The film exhibited a uniform contrast, indicating a consistent monolayer thickness. We note that the line features with bright contrast, seen in Figure 3b, are wrinkles and are not associated with grain boundaries (Figure S8).

The uniform growth of monolayer hBN with aligned crystallinity is scalable, resulting in a continuous film over a 2-inch Ge(110) substrate (Figure 3c, Figure S9). The hBN film transferred onto fused silica (Figure 3c, inset) showed similar optical transmission contrast spectra, 1-$T$ at each measurement point across the entire film, which are consistent with those for monolayer hBN.[20] Here, $T$ is defined as $I/I_0$, with $I_0$ and $I$ representing the transmission intensity through the bare substrate and the intensity through the substrate with the film, respectively.

To further characterize the hBN film, we conducted optical second harmonic generation (SHG) measurements. The SHG signals were measured from the hBN film transferred onto a SiO$_2$/Si substrate in reflection mode using an excitation laser with a fundamental wavelength of 800 nm. The polarization-dependent SHG plot (Figure 3d) was measured with a parallel-aligned linear polarizer and analyzer while changing the relative orientation, $\Phi$ of the polarization with respect to the sample. The SHG intensities show a 6-fold symmetry, fitted as $\cos^2(3\Phi)$, representing the aligned crystallographic orientation of hBN.[21] Additionally, the SHG map (Figure 3e) shows uniform intensity, suggesting a lack of grain boundaries, where SHG is suppressed by destructive phase interference,[22] and multilayer regions with different stacking orders.[23]

## 2.4. DFT calculations of binding energy between hBN and Ge

We conducted DFT calculations to understand how the hBN grains are aligned with respect to the Ge epi-layer (**Figure 4**a). Unlike metallic substrates for epitaxial hBN growth,[8-14] the Ge surface might be passivated by hydrogen atoms. At high $P_{H2}$, both hBN edges and Ge surface are expected to be hydrogen-passivated (Figure 4b). Our DFT calculations show that, at the edge of hBN, the nitrogen-hydrogen (N-H) bonding strength is 5.34 eV per bond, which is significantly higher than that of Ge-hydrogen bonds (Ge-H) on Ge(110) surface, 3.11 eV per bond. Therefore, hBN edges are more likely to be hydrogen-passivated than the Ge surface. To understand the epitaxial growth behavior of hBN on Ge(110), we calculated the binding energies ($E_B$) of a hBN island on both Ge(110) surface with and without hydrogen passivation. The calculated binding energies of hBN with different orientation angles are shown in Figure 4c, from which we can clearly see two local minima at $\theta = 0°$ and $60°$ for both curves. For each case, the $E_B$ at $\theta = 0°$ and $60°$ are very close to each other. The tiny energy difference of less than 0.01 eV made the orientations of $\theta = 0°$ and $60°$ undistinguishable, which explains that the hBN islands grown on Ge(110) terrace generally have two orientations. On the other hand, the energy barriers of rotating the hBN island on pristine Ge(110) surface is approximately 0.78 eV, which is more than one order of magnitude larger than that on the hydrogen-passivated Ge(110) surface, which is only about 0.046 eV.

It is known that the introducing atomic steps on a flat terrace may lead to the uni-directional alignments of 2D islands due to the strong interaction between the step edge and the hBN edge. So, we further considered the binding energy between the hBN edge and the step edge of the Ge(110) surface. At a high hydrogen pressure, both the hBN edge and the whole Ge(110) surface are terminated by hydrogen and, therefore, the interaction between the hydrogen-terminated hBN edge and the hydrogen-passivated Ge(110) surface is of the weak van der Waals-type. As shown in Figure 4d, the weak step edge-hBN edge interaction only leads to a very shallow energy minimum in the $E_B$ vs. $\theta$ curve at $\theta = 0°$ and $E_B$ at $\theta = 60°$ is a local minimum. Consequently, both hBN domains with $\theta = 0°$ and $60°$ are of very similar stabilities. This result is in agreement with our observations of hBN bi-crystal grown at high $P_{H2}$.

As shown in Figure S10, at a reduced hydrogen pressure (or a lower chemical potential of hydrogen), direct chemical bonds between the Ge step edge and the hBN edge (as shown in Figure 4e) are energetically favorable. We calculated the formation energy of the Ge step edge-hBN interface as a function of $\theta$ and the result is shown in Figure 4f, from which we can see that $\theta = 0°$ corresponds a deep global minimum of the formation energy curve and $\theta = 60°$ corresponds a global maximum. Therefore, at an intermediate $P_{H2}$, which is sufficient to passivate both the Ge substrate and the hBN edge but also allow the formation of Ge-N bonds between the hBN edge and the Ge step edge, step-directed uni-directional growth of hBN can be achieved. The simulation results are consistent with the experimentally observed non-monotonic change of $\chi_{align}$ as a function of $P_{H2}$.

**2.5. Applications of single-crystalline hBN films**

The fact that Ge surfaces can be hydrogen-passivated makes them unique compared to other epitaxial metal substrates used for growing uni-directional hBN.[18] Due to the hydrogen-passivation, the hBN films can be mechanically exfoliated and assembled to form atomically clean vdW interfaces without exposure to other chemicals, as shown in **Figure 5**a. Angle-resolved photoemission spectroscopy (ARPES) data, conducted on a few-hundred-micrometer-wide region of the as-grown monolayer hBN film, reveal clear bands expected for hBN with aligned crystallography (see Figure 5b). Using these films, we have fabricated an hBN bilayer by layer-by-layer assembly with aligned crystallography. The ARPES data (Figure 5c) demonstrate clear band splitting along the Γ-K momentum direction attributed to interlayer $P_z$ orbital hybridization,[24] confirming that the fabricated interfaces are pristine. Since the hBN assembly units possess uni-directional crystallinity, we can further control the stacking order of the hBN bilayers, which significantly affects their electrical and optical properties.[25-29] For demonstration, we assembled two monolayer hBN films with the crystallographic alignment either parallel or anti-parallel. The diffraction patterns from transmission electron microscopy reveal a single hexagonal pattern for the bilayer samples with different stacking orders, indicating precise alignment (Figure 5d and e). However, the relative intensity ratios between the inner and outer spots differ, as shown in the insets of Figures 5d and e, which are consistent with targeted rhombohedral and hexagonal stacking orders, respectively.[23] Although a slight misorientation angle from the targeted alignment (<1°) may occur during assembly, atomic reconstruction tends to yield stable atomic configurations with overlapping elements of opposite charge polarities (i.e., boron on nitrogen, or nitrogen on boron).[23] We note that the rhombohedral stacking order, with its distinct crystal symmetry (non-centrosymmetric, lacking a mirror plane parallel to the layer), results in emergent properties such as ferroelectricity and nonlinear optical characteristics. Despite significant interest, producing this metastable phase on a large-scale has been challenging. Our layer-by-layer assembly of uni-directionally grown films now provides a method for fabricating this film.

We also fabricated graphene (Gr) and $MoS_2$-based field-effect transistors (FETs) using the hBN film as a dielectric interlayer (Figure 5f). (see Method for the details) Electrical transport in the Gr FETs was compared between hBN and $SiO_2$ dielectric. In the sheet resistance ($R_\square$) versus applied gate, $V_{BG}$ curves, measured with a sweep rate for field-induced carrier density of $3\times10^{10}$ cm$^{-2}\cdot$s$^{-1}$ (Figure 5g), we observed significant suppression of hysteresis with the hBN dielectric, implying effectively suppressed interfacial traps. To quantify the interfacial trap densities, we deduced the change of surface carrier concentration, $\Delta n$, induced by $V_{BG}$. Here, $\Delta n = V_{BG}C_g/e$, where $C_g$ is the gate capacitance per unit area, and $e$ is the elementary charge. The difference in $\Delta n$ at the charge neutrality points for the forward and reverse sweeps indicates the trap density. The reduced density is lowered to $3.50\times10^{11}$ cm$^{-2}$ with hBN dielectric from $3.48\times10^{12}$ cm$^{-2}$ with $SiO_2$ dielectric, representing a reduction by roughly an order of magnitude. The graphene field-effect mobility, $\mu$, was also extracted by fitting the transconductance curves with $\mu = dG_\square/d\Delta n \cdot (1/e)$ based on Drude model, where $G_\square$ indicates sheet conductance. As a result, $\mu$ for electrons and holes are 108 and 1394 cm²/V·s for SiO$_2$ dielectrics, and 207 and 723 cm²/V·s for hBN dielectrics. We suspect that Coulomb interactions through the ultrathin hBN dielectric of ~1.4 nm between the channel and the gate affected the field-effect mobility. Nonetheless, the significant suppression of hysteresis in the transconductance sweep was evident, indicating a reduction of trap states in the dielectric compared to silicon oxides [30].

In MoS$_2$ FET, we compared the subthreshold swing (SS) and mobility ($\mu$) from transconductance curves with and without the hBN interlayer on a SiO$_2$/Si substrate (Figure 5h). The introduction of hBN increases $\mu$ from 12 cm$^2 \cdot$V$^{-1} \cdot$s$^{-1}$ to 15.8 cm$^2 \cdot$V$^{-1} \cdot$s$^{-1}$ and decreases the SS from 2.3 V/dec to 1.2 V/dec. The SS can be expressed as $\ln(10) \cdot V_{th} \cdot \{1+(C_{ch}+e \cdot D_{it})/C_{ins}\}$, where $V_{th}$ is the thermal voltage, $C_{ch}$ is the channel capacitance, $D_{it}$ is the density of interface states, and $C_{ins}$ is the insulator capacitance.[31] The difference in the deduced $D_{it}$ with and without hBN is estimated to be 8.27 × 10$^{11}$ cm$^{-2}$ eV$^{-1}$.

## 3. Conclusion

In conclusion, we have grown large-scale hBN monolayer films with uni-directional crystallography on miscut Ge(110) substrates using CVD. Our experimental and theoretical studies suggest that Ge step-assisted epitaxy thermodynamically determines the orientation of hBN grains when the $P_{H2}$ during growth is optimized to passivate the Ge terrace, while still allowing direct bonding between hBN grains and the Ge steps at the nucleation stages. For the first time, the uni-directional hBN monolayer films were mechanically exfoliated from the growth substrates for atomically clean assembly, implying that this process can be useful for fabricating multilayered hBN with controlled stacking orders, and for creating advanced electronic and optoelectronic devices using vdW materials.[32-34]

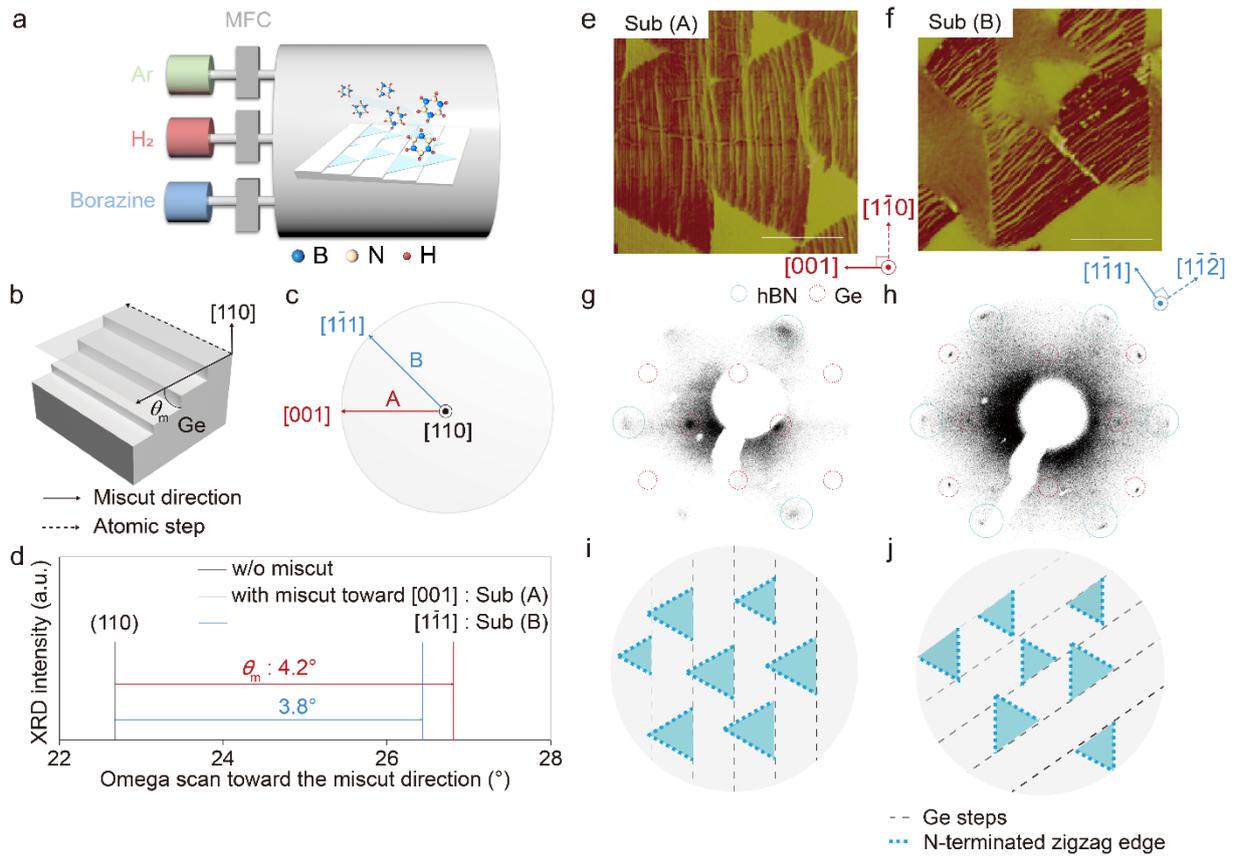

**Figure 1. Epitaxial relationship between hBN and atomic steps of Ge.**

(a,b) Schematic of the hBN growth setup and a miscut Ge(110) wafer, where the surface is tilted toward the miscut direction by angle $\theta_m$ with respect to the (110) plane. (c) Top view of a miscut Ge(110) wafer with miscut directions [001] for Sub(A) and [11$\bar{1}$] for Sub(B). (d) XRD rocking curve for Ge(110) wafers without miscut and with miscut toward different directions. (e, f) AFM phase images of partially grown monolayer hBN on Sub(A) and Sub(B), respectively. (Scale bar: 1 μm) (g, h) LEED patterns of the samples on Sub(A) and Sub(B) over an area of 1 mm². (i, j) Schematics of partially grown hBN on miscut Ge(110). Two hBN edges not aligned with the Ge steps (marked as dotted black lines) are N-terminated zigzag edges (indicated by dotted sky blue lines) on both Sub(A) and Sub(B).

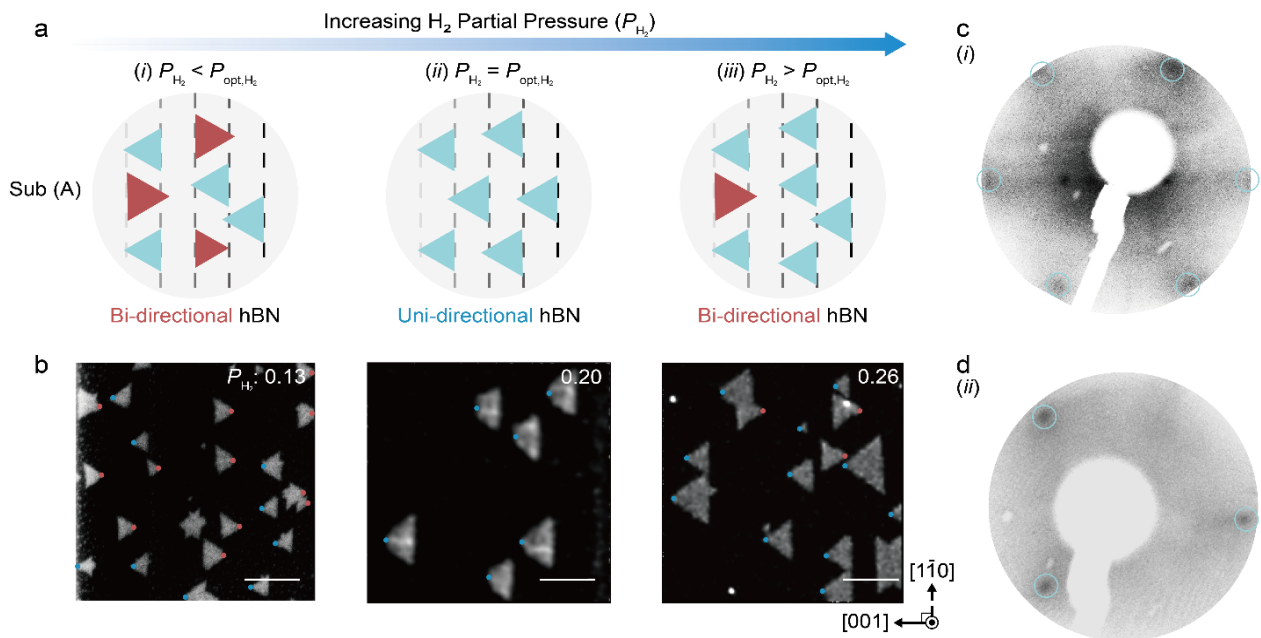

**Figure 2. Effect of hydrogen partial pressure on the crystallographic alignment of hBN.**
(a) Schematic illustration of the variation in the degree of hBN alignment as $P_{H2}$ increases. (b) SEM images of hBN grains with increasing $P_{H2}$. (Scale bar: 1 μm) (c, d) LEED patterns of (c) hBN film with bi-directional crystal growth and (d) uni-directional crystal growth.

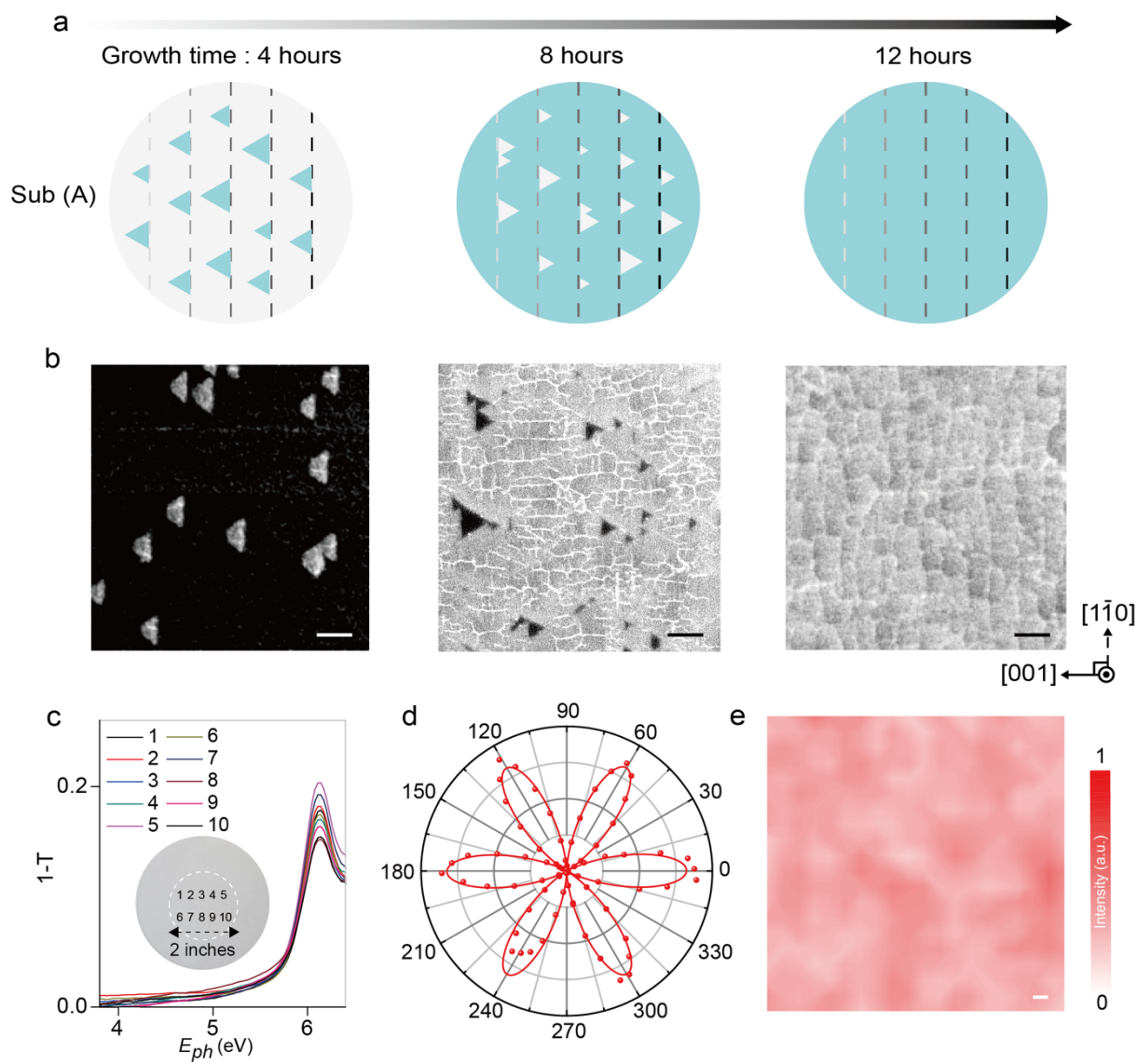

**Figure 3. Wafer-scale formations of single-crystalline hBN.**

(a) Schematic of the sequential process of hBN film formation through the merging of uni-directional grains on Ge(110) substrates with a miscut. (b) SEM images of as-grown hBN samples at different growth times (4, 8, and 12 hours from left to right). (c) Optical transmission contrast spectra measured at the positions indicated in inset (inset : the hBN film transferred onto fused silica, with the hBN film region highlighted by a dotted line). (d) Polar plot of the SHG intensity as a function of the crystal's azimuthal angle $\Phi$. (e) SHG intensity map. Scale bar: 1 μm.

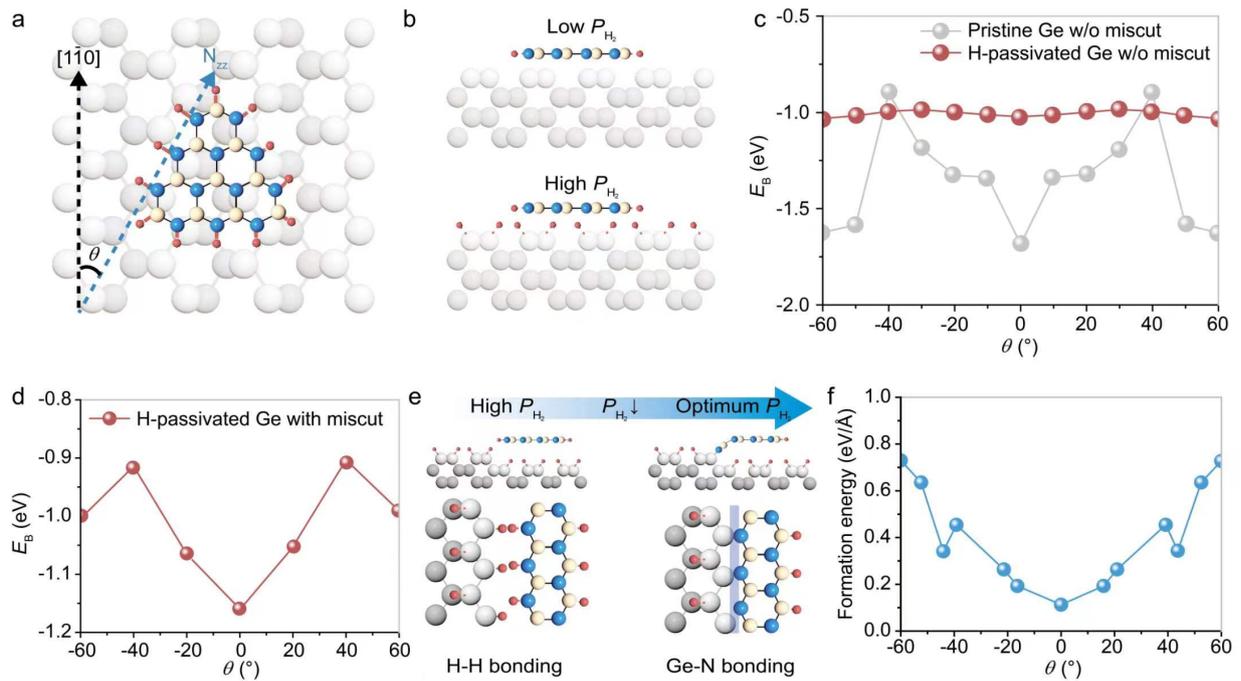

**Figure 4. DFT calculations of binding energy between hBN and Ge.**
(a) Schematics of an hBN grain with N-terminated zigzag edges on Ge(110). The angle ($\theta$) indicates the misorientation between the hBN edge and the Ge [1$\bar{1}$0] direction. (b) Schematics of the side view of hBN on (top) pristine Ge(110) and (bottom) hydrogen-passivated Ge(110). (c) Binding energy between hBN and Ge(110) (gray) without hydrogen passivation and (red) with hydrogen passivation as a function of $\theta$. (d) Binding energy between hBN and hydrogen-passivated Ge(110) with a miscut as a function of $\theta$. (e) Schematic illustration of bond formation between a Ge atomic step and hBN at different $P_{H2}$. (f) Formation energy for one-dimensional hBN/Ge step interfaces as a function of $\theta$, when hBN and Ge form direct chemical bonds without hydrogen passivation.

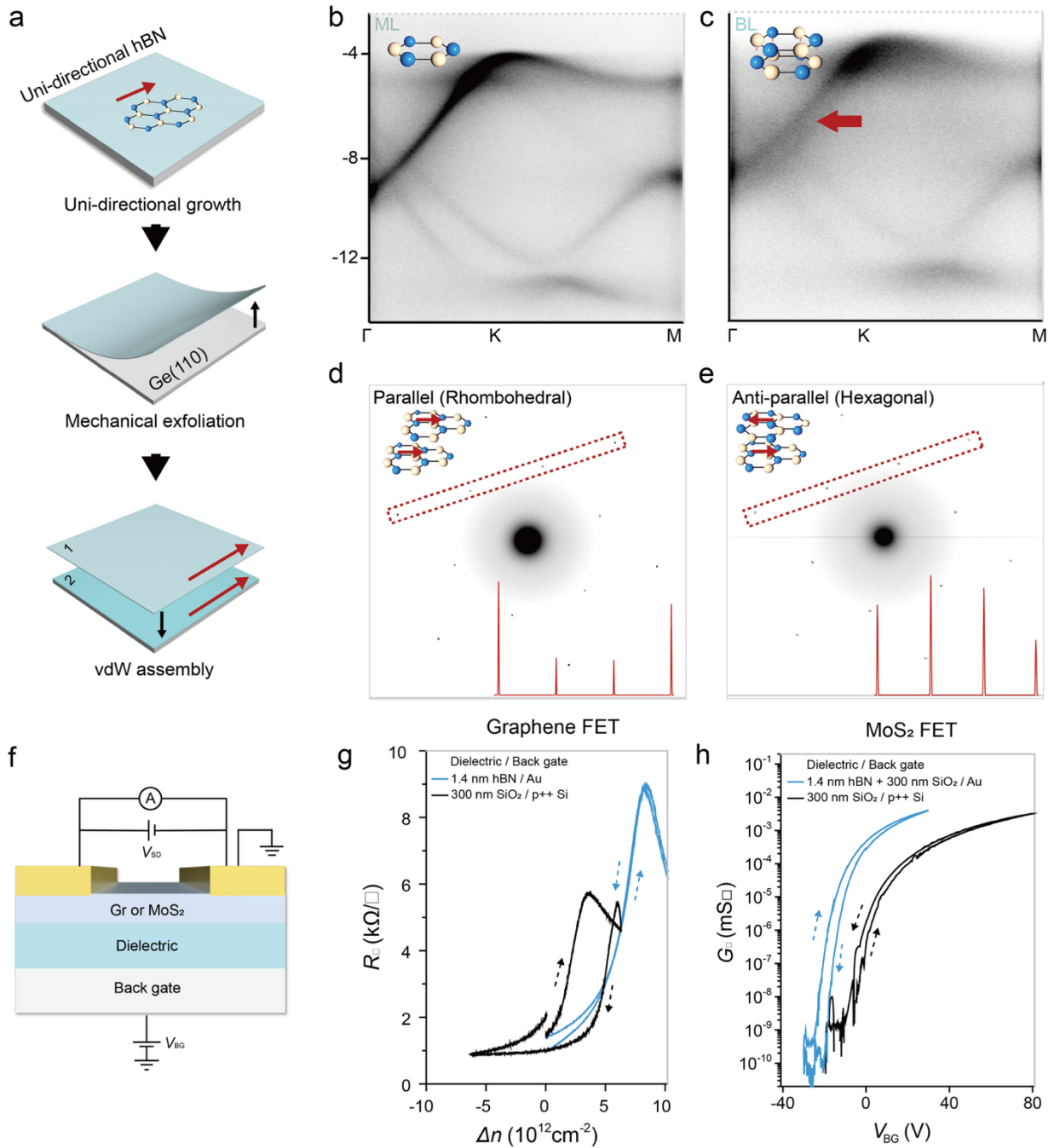

**Figure 5. Applications of single-crystalline hBN films.**
(a) Schematic illustration of the layer-by-layer assembly process. (b, c) ARPES data for (b) as-grown monolayer and (c) assembled bilayer hBN films. In the assembled bilayer hBN, band splitting (indicated by a red arrow) occurs along the Γ-K direction, representing effective interlayer coupling. (d, e) TEM diffraction patterns of bilayer hBN assembled with (d) parallel and (e) anti-parallel interlayer alignment of crystallography. (f) Schematic of Gr and $MoS_2$ FET devices. (g, h) (g) Sheet resistance versus $V_{BG}$ for Gr FETs, and (h) sheet conductance versus applied gate for $MoS_2$ FETs with different dielectrics.

## 4. Experimental Section

**Growth of hBN:** Ge (110) substrates miscut by 4° toward [001] or [1$\bar{1}$0] (PAM-XIAMEN) were cleaned in 20% HF solution for 3 min to remove native oxides. The Ge substrates were then loaded in a quartz tube, which was evacuated to 40 mTorr and subsequently refilled to atmospheric pressure with combinations of 99.99% $H_2$ and 99.99% Ar. The tube was annealed until it reached 930 °C over 1 hour, with various ratios of $H_2$ to Ar (At the optimum hydrogen partial pressure for uni-directional growth, 4000 sccm of Ar and 1000 sccm of $H_2$ were used). Borazine (Gelest-INBO009) was then introduced for 12 h using Ar as a carrier gas (0.3 sccm) to grow a uni-directional monolayer hBN film. After the growth of hBN, the borazine flow was turned off and the furnace was opened to cool naturally.

**Characterizations:** All LEED patterns were obtained over an area of ~1 $mm^2$, using an optics system (ErLEED 150) at 98 eV and $10^{-9}$ Torr. XRD patterns of vicinal Ge substrates were acquired using an X-ray diffractometer (Empyrean, Panalytical) with Cu Kα radiation. The hBN domains were observed using a high-resolution field-emission SEM (Hitachi S-4800) by detecting secondary electrons from hBN/Ge surface. Band structures of hBN were investigated at the 10D beamline at the Pohang Accelerator Laboratory, using an electron analyzer (Scienta DA30) with a photon energy of 55 eV. Additionally, hBN films were transferred onto a SiN TEM grid (EMS, 76042-48) for diffraction pattern analysis and dark field imaging at 200 keV using a transmission electron microscope (Jeol, JEM-2100F).

**Layer-by-layer assembly:** An Au layer was deposited to 40-nm thickness on a hBN/Ge(110) sample. Then, polymethyl methacrylate (PMMA, 996 K, 8% in anisole) was spin-coated onto the Au film at 1500 rpm for 15 s, then annealed at 190 °C for 15 min on a hot plate. A thermal-release tape (TRT, Nitto 3195HS) was attached to the top of the PMMA/Au/hBN/Ge(110) sample as a handling layer, to enable mechanical exfoliation of TRT/PMMA/Au/hBN. Then the exfoliated TRT/PMMA/Au/hBN film was attached to another as-grown hBN on Ge(110) sample with controlled stacking orientation and baked at 180 °C for 10 min to increase the mechanical-exfoliation yield. By repeating the assembly process with different twist angles, we can control the stacking order and the total number of layers in the final films (Figure S11a). Finally, the Au film was etched with KI/$I_2$ solution, and the surface was rinsed with ethanol and deionized (DI) water. The pristine interfaces of the assembled hBN multilayers were confirmed by XRD (Figure S11b). Please, refer to our previous report[18] for more details.

**Computational details:** All the DFT calculations are performed using the Vienna *Ab initio* Simulation Package (VASP).[35, 36] The generalized gradient approximation and projected augmented wave method were used to describe the exchange-correlation functions and the interaction between electrons and ion cores. The plane wave cut-off energy was set as 300 eV, and 1 × 1 × 1 *k*-point meshes were adopted for these DFT calculations. The convergence criteria of the forces and energies were set to be 0.02 eV·Å$^{-1}$ and 1.0 × 10$^{-4}$ eV, respectively. A vacuum layer with a thickness greater than 15 Å was used to avoid the interactions between two neighboring images. The vdW interactions were corrected by the DFT-D3.[37] We constructed a model of monolayer hBN domain on a three-layer Ge substrates with the bottom layer fixed and hydrogen- passivated. The binding energy ($E_B$) and formation energy ($E_f$) are calculated using the following two equations:

$$E_B = (E_{total} - E_{Ge} - E_{hbn})/A, \quad (1)$$

$$E_f = (E_{total} - E_{Ge} - E_{hbn} + \varepsilon_{pri})/L, \quad (2)$$

where $E_{total}$, $E_{Ge}$ and $E_{hbn}$ are the energies of the whole system, the Ge substrate and the hBN layer, respectively. *A* is the area of the hBN domain. *L* is the length of the supercell of the hBN/Ge(110) system. $\varepsilon_{pri}$ is the formation energy of hBN with a free edge, which is referred to our previous study.[10] To compare the free-energy difference between Ge-H and the Ge-N edges, two atomic configurations are modeled: both hBN and a stepped Ge(110) surface with H-passivated (Figure S10, *i*) and the model with one H removed from hBN and Ge(110) surface (Figure S10, *ii*), respectively. The free energy for the aforementioned models can be calculated using the following equations:

$$G_i = (E_{total} - E_{Ge} - E_{hbn}) - 0.5n_H(E_{H_2} + \Delta\mu_{H_2}), \tag{3}$$

$$G_{ii} = (E_{total} - E_{Ge} - E_{hbn}) - 0.5(n_H - 2)(E_{H_2} + \Delta\mu_{H_2}), \tag{4}$$

where $E_{total}$, $E_{Ge}$, $E_{hbn}$, $n_H$, $E_{H_2}$ and $\Delta\mu_{H_2}$ represent the total energy of the supercell, the energy of the Ge substrate, the energy of the hBN domain, the number of hydrogen atoms at the hBN edge, and the energy of a hydrogen molecule (-6.7617 eV/molecule), and the chemical potential of hydrogen, respectively. The free energies as a function of $H_2$ chemical potential for the aforementioned models were further calculated in Figure S10.

**Device fabrications:** 4 to 5-layer hBN films were first fabricated onto a Ge substrate using a layer-by-layer assembly method. This assembled 4 layers of hBN was subsequently transferred onto a Si substrate for Gr FETs, and 5 layers was transferred onto a $SiO_2$/Si substrate for $MoS_2$ FETs. Then, Gr or $MoS_2$ was transferred onto the hBN films. Metal electrodes, Cr/Au (15/40 nm) for Gr FETs and Bi/Au (15/40 nm) for $MoS_2$ FETs, were deposited using a thermal evaporator and patterned by photolithography.

**Author Contributions**

J.-H. Jung and C. Zhao contributed equally to this work. J.-H. Jung and C.-J. Kim designed the experiments. J.-H. Jung synthesized the hBN samples. C. Zhao and F. Ding performed the DFT calculations. J.-H. Jung, S.-J. Yang, J.-H. Park, and C.-C. Hwang conducted the structural characterizations on hBN, including TEM and LEED. S.-J. Yang and W.-J. Lee fabricated and characterized the electronic devices. S.-B. Song and J. Kim conducted optical second harmonic generation and analyses. S.-H. Baek conducted XPS and FTIR analyses on hBN. J.-H. Jung and C.-J. Kim wrote the manuscript with input from all authors.

## Acknowledgements

This research was supported by the National R&D Program through the National Research Foundation of Korea (NRF) funded by the Ministry of Science and ICT (RS-2023-00234622, 2023R1A2C2005427, RS-2023-00258309) and the Institute for Basic Science (IBS-R034-D1).

# Supporting Information

| hBN fabrication | Lateral size | Thickness controllability | Atomically clean assembly | Single-crystallinity | Ref. |
|---|---|---|---|---|---|
| Exfoliation from bulk crystals | tens of μm$^2$ | X | O | O | [1], [2] |
| Growth of multilayer films on metallic substrates | > cm$^2$ | O | X | O | [3], [4], [5] |
| Layer-by-layer assembly of monolayer hBN | > cm$^2$ | O | O | X | [6] |
| | | O | O | O | This study |

**Table S1**. Summary of different hBN fabrication methods.

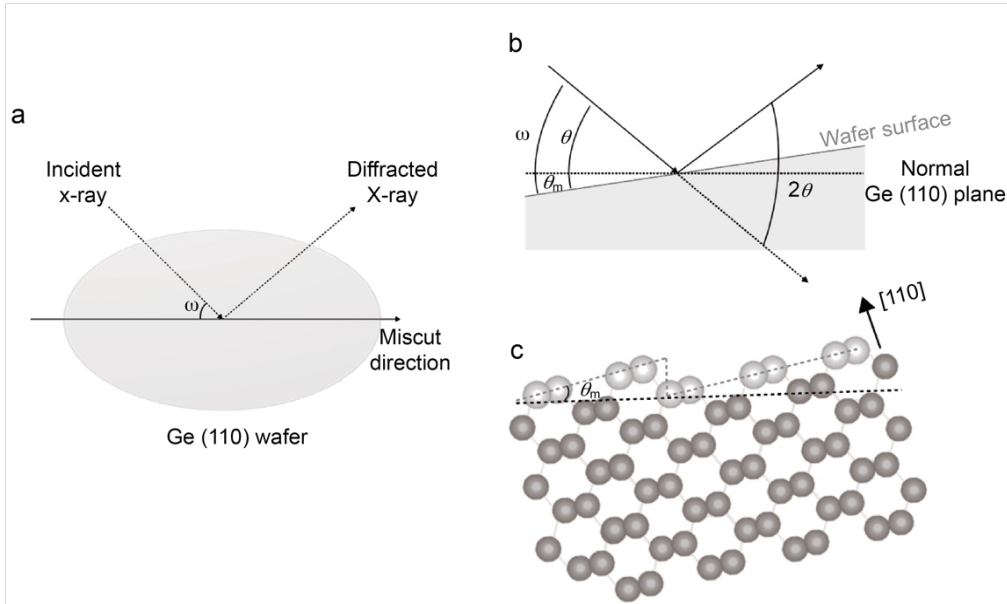

**Figure S1. Measurement of miscut angle of Ge(110) wafer through XRD.**

(a) Schematics of incident and diffracted X-ray parallel to miscut direction of Ge(110). (b) The omega ($\omega$) is the angle between incident X-ray and the Ge surface, when (110) plane is situated flat. As $\omega$ is the sum of diffraction angle $\theta$ and miscut angle $\theta_m$, one can deduce the $\theta_m$. (c) Side view of atomic structure of miscut Ge(110).

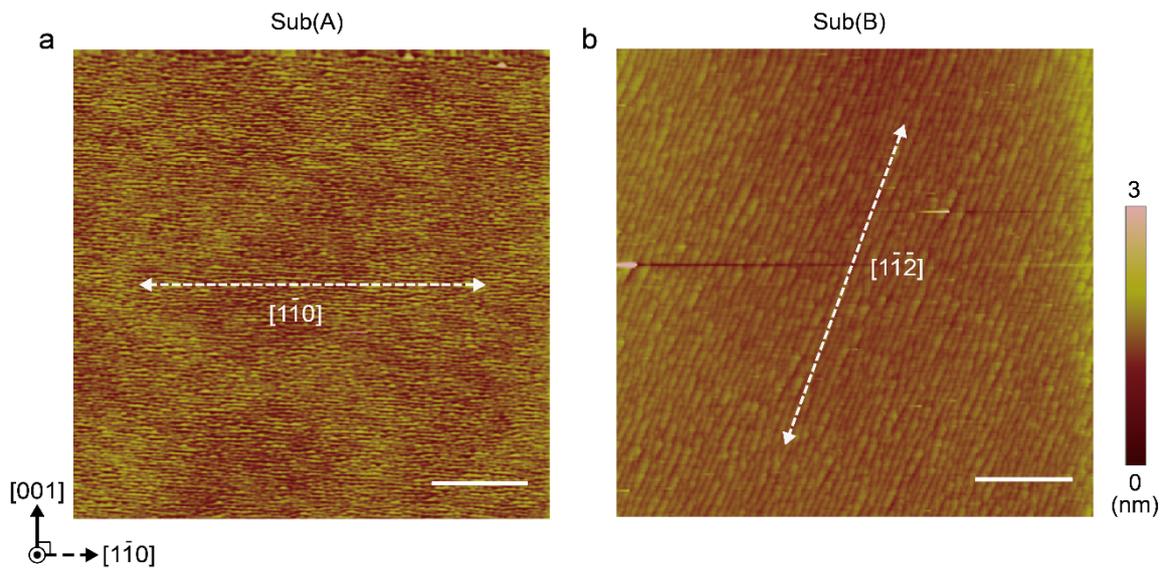

**Figure S2. Atomic step directions of miscut Ge(110) substrates.**

(a, b) AFM height images on (a) Sub(A) and (b) Sub(B). (Scale bar: 1 μm)

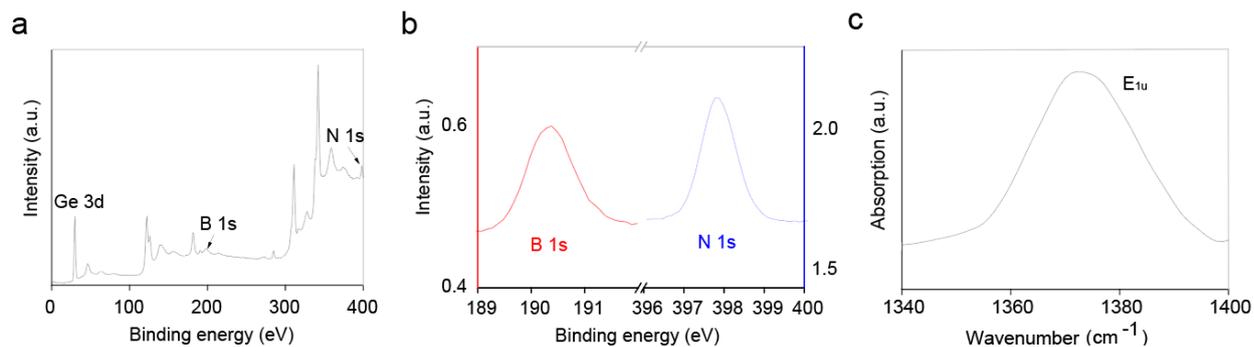

**Figure S3**. Spectroscopy analyses of as-grown hBN on Ge (110). (a, b) XPS spectra. The presence of boron (B) and nitrogen (N) elements was confirmed with distinct peaks for B 1s and N 1s observed at 190.3 eV and 397.8 eV, respectively. The atomic ratio of B to N in the sample is estimated to be 1.02:1, considering the atomic factor. (c) FTIR spectra. A clear peak in FTIR appears at 1372 cm$^{-1}$, matching the infrared-active E1u mode for stretching in-plane B-N bonds in hexagonal boron nitride.

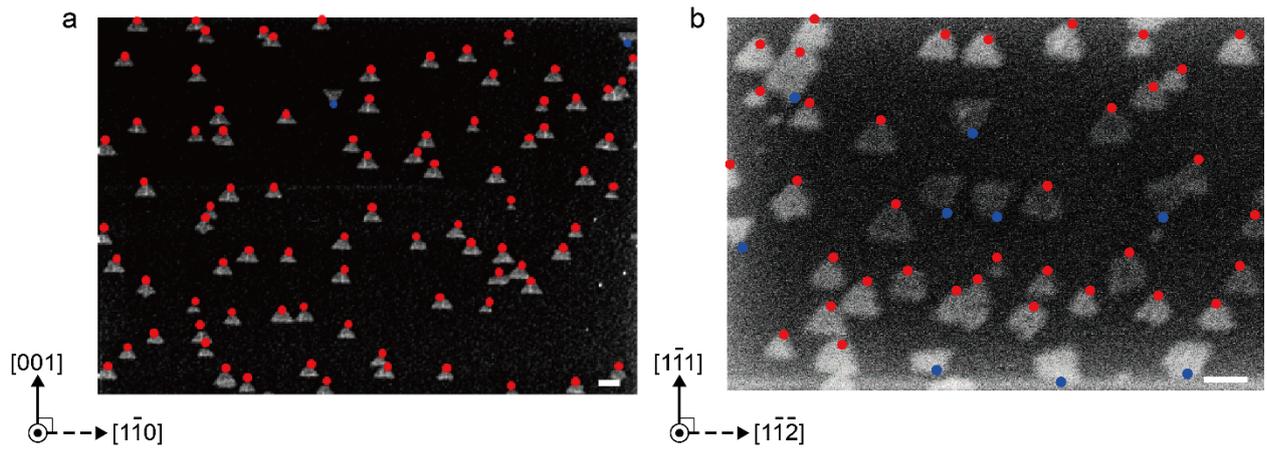

**Figure S4. Comparison of the orientation of hBN grains grown on Sub(A) and Sub(B).**
(a, b) Representative SEM images of monolayer hBN domains with partial coverages grown on (a) Sub(A) and (b) Sub(B). The grains are classified into two group according to the direction of apex of triangular grains from the axis of Ge steps (dotted lines in the bottom left corner); the grains are highlighted with red dots when they are pointing to the miscut direction parallelly, and with blue dots when pointing to the miscut direction anti-parallelly. The proportion of red-highlighted grains is 97 % on Sub(A) and 77 % on Sub(B). (Scale bar: 1 μm)

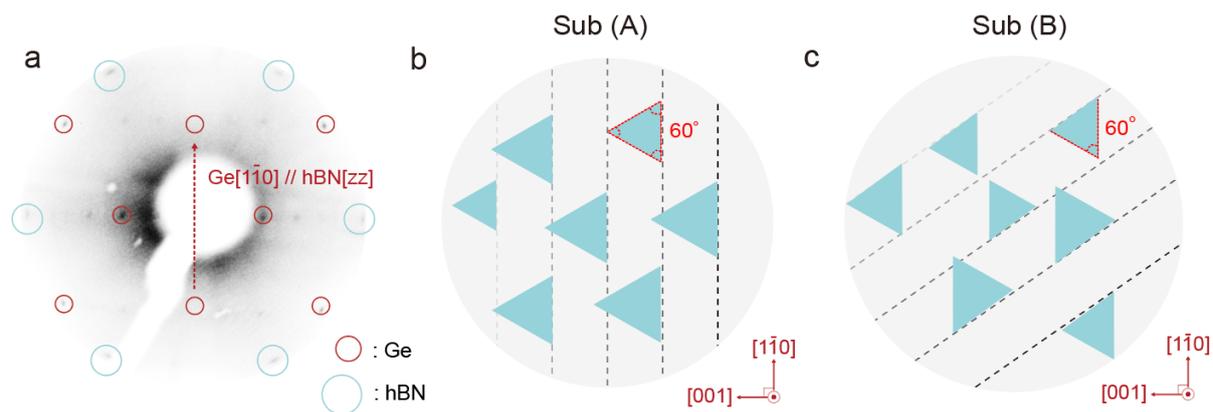

**Figure S5.** (a) The LEED pattern of hBN on Ge (110). (b,c) Schematic of the angles between hBN edges on Sub(A) and Sub(B).

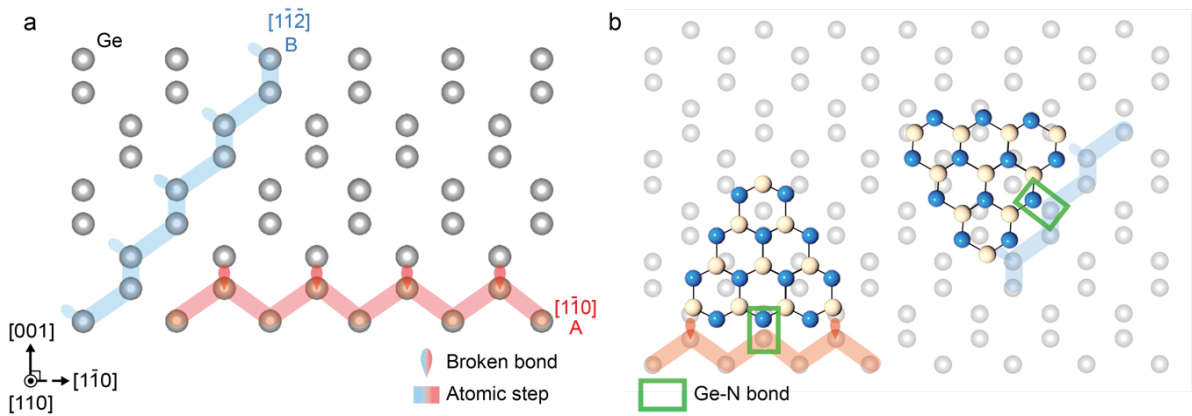

**Figure S6. Ge-N bonding density at atomic step on Sub(A) and Sub(B).**

(a) Schematics for dangling bonds of Ge on Sub(A) and Sub(B) with different orientations for atomic steps. (b) Schematics of atomic configuration for hBN grains attached to each Ge atomic step. The dangling bonds along the Ge step are assumed to be fully passivated with hBN, as their density is lower than the density of N atoms along the hBN edge. With the assumption, the Ge-N bond densities are estimated as 0.25 nm$^{-1}$ for Sub(A), but 0.21 nm$^{-1}$ for Sub(B).

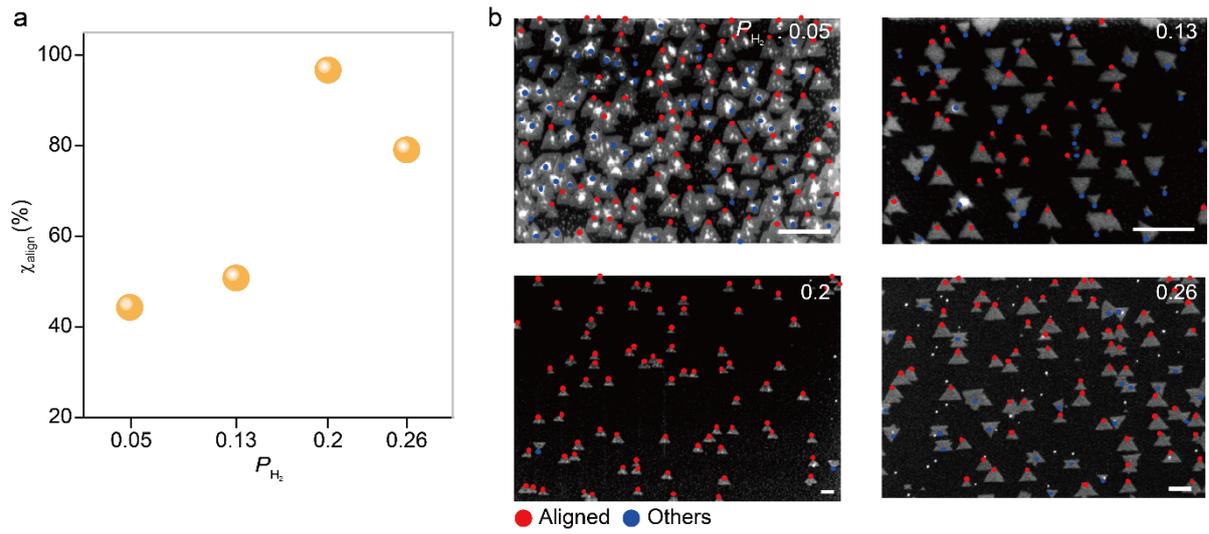

**Figure S7. Orientation of hBN grains dependent on hydrogen partial pressure.**

(a) $\chi_{align}$ as a function of $P_{H2}$ (refer to the main manuscript for the definition). (b) Representative SEM images of hBN grains partially grown on Sub(A) under various $P_{H2}$. (Scale bar: 1 μm.)

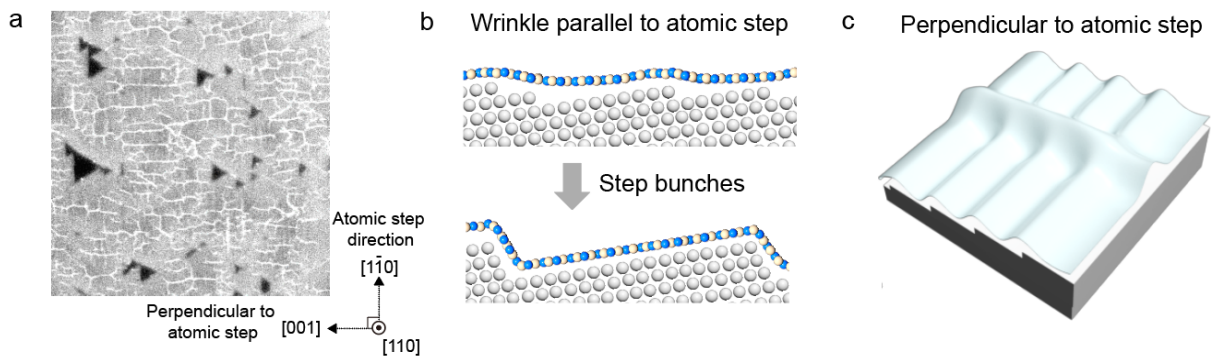

**Figure S8. Dynamics of wrinkle formation in hBN on vicinal Ge (110).**

(a) Alignment of hBN wrinkles relative to atomic steps. (b) Wrinkle formation parallel to atomic steps during cooling. (c) Wrinkle formation induced by de-adhesion of hBN. The line features with bright contrast in (a) are mostly aligned with the horizontal axis in the images, perpendicular to the atomic steps. Recent studies[7-9] suggest that the bunching of atomic steps in the substrate during growth near the melting point of the substrate forms larger terraces with low-index planes of lower surface energy, increasing the step heights, which form slightly corrugated structures in the grown films, as seen in (b). Stress is then applied to the films during cooling due to the mismatch in thermal compression with the growth surfaces. While the stress along the corrugated direction can be effectively relaxed, the relaxation of the stress along the atomic steps is achieved only by the de-adhesion of hBN, resulting in wrinkles perpendicular to the atomic steps, as depicted in (c).

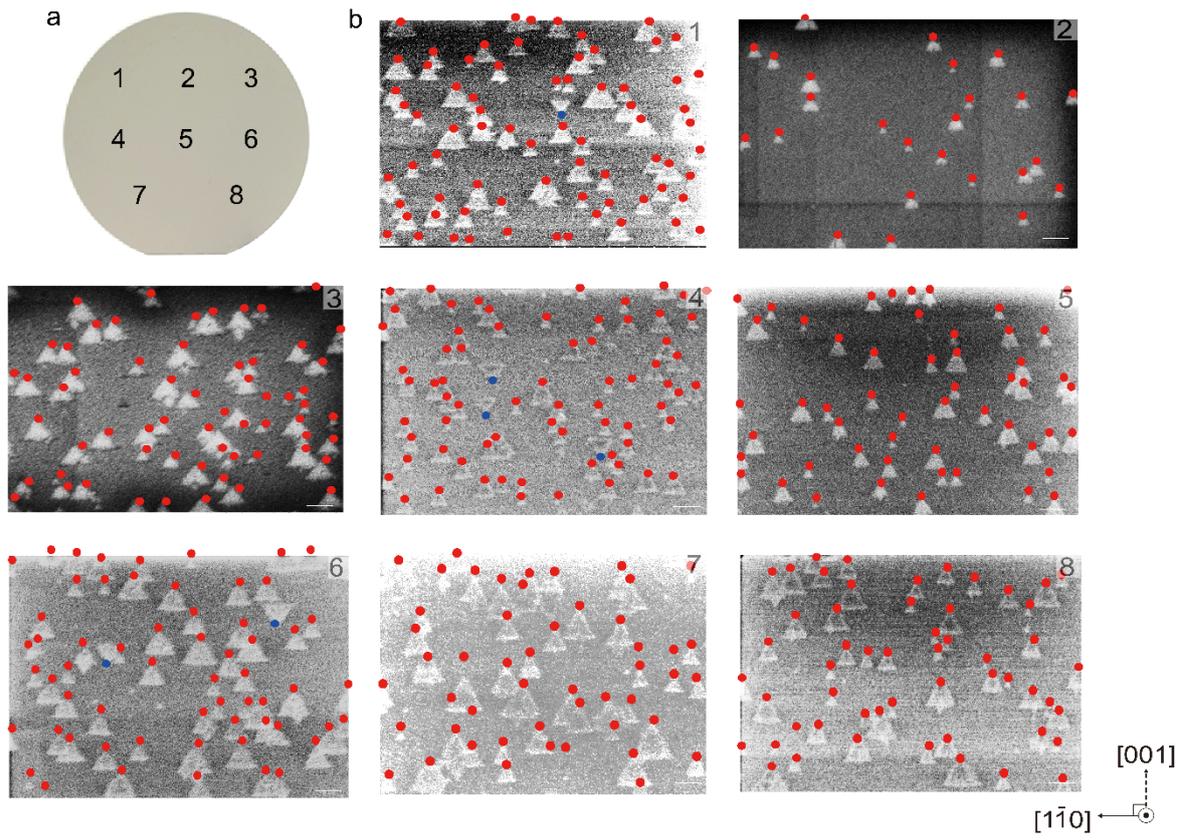

**Figure S9.** (a) A photograph of 2-inch Ge(110) wafer. (b) SEM images measured at the positions indicated in (a). Scale bar: 1 μm.

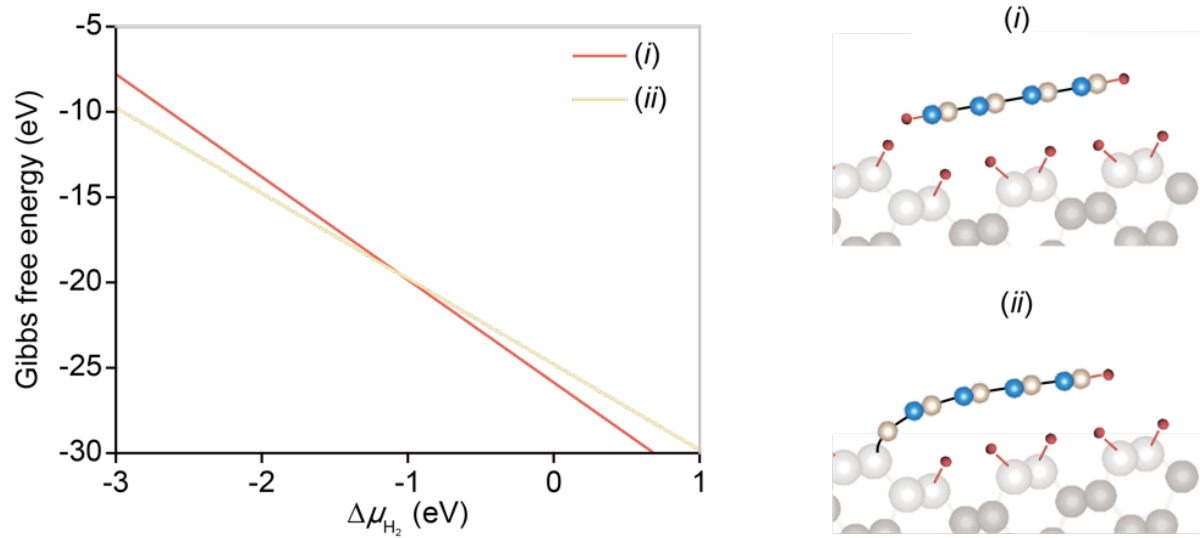

**Figure S10. Gibbs free energy of an hBN grain attached to a Ge step.**

Gibbs free energy of an hBN grain attached to a Ge step as a function of hydrogen chemical potential, considering both (*i*) H-H bonding and (*ii*) Ge-N bonding cases. At lower $\Delta\mu_{H_2}$, formations of direct Ge-N bonds are preferred.

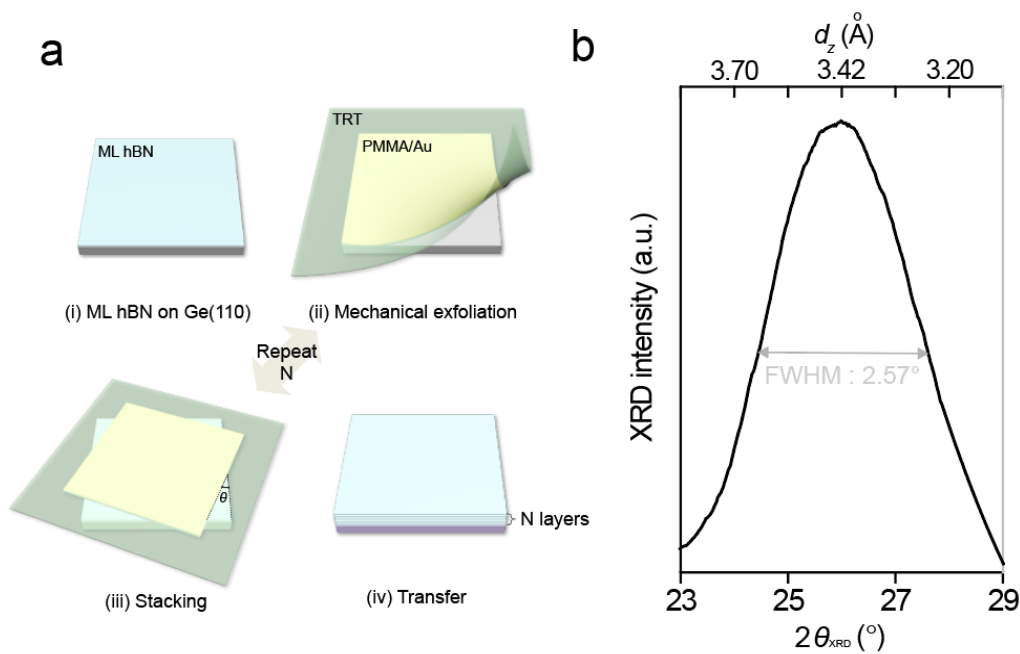

**Figure S11.** (a) Schematic of mechanical layer-by-layer assembly process of hBN films. (b) XRD spectra of assembled 10 hBN layers on Ge(110). The interlayer distance, deduced from the peak position, was estimated to be approximately 3.42 Å, which closely matches the interlayer distance of bulk hBN crystals. The crystallographic orientations of assembled hBN layers were not aligned, which may result in a slightly larger interlayer distance than that of bulk crystals due to lattice distortion[10]. Additionally, the thickness of coherent lattices was estimated using the Scherrer equation, represented as $K\cdot\lambda/(\beta\cdot\cos(\theta_{XRD}))$, where $K$ is a shape factor (0.9), $\lambda$ is the wavelength of Cu $K_\alpha$ radiation (1.5406 Å), $\beta$ is the full width at half maximum of the $2\theta_{XRD}$, and $\theta_{XRD}$ is the Bragg peak position. The estimated value was 3.17 nm, which closely matches the thickness of the stacks of 3.4 nm, indicating the formation of pristine interfaces that facilitate constructive interferences.